\documentclass[a4paper]{article}

\usepackage{INTERSPEECH2020}
\usepackage[export]{adjustbox}
\usepackage{subfigure}
\usepackage{hyperref}
\usepackage[shortlabels]{enumitem}
\usepackage{CJKutf8}

\usepackage{color}
\usepackage{algorithm}  
\usepackage{algpseudocode}  
\usepackage{amsmath}  

\hypersetup{
    colorlinks=false,
    linkcolor=blue,
    filecolor=blue,      
    urlcolor=blue,
    citecolor=cyan,
}

\title{Glow-WaveGAN: Learning Speech Representations from GAN-based Variational Auto-Encoder For High Fidelity Flow-based Speech Synthesis}
\name{Jian Cong$^{1}$\thanks{Work performed when Jian Cong was interning at Tencent AI Lab. Lei Xie is the corresponding author.}, Shan Yang$^2$, Lei Xie$^1$, Dan Su$^2$}
\address{
  $^1$Audio, Speech and Language Processing Group (ASLP@NPU), School of Computer Science, Northwestern Polytechnical University, Xi'an, China\\
  $^2$ Tencent AI Lab, China}
\email{npujcong@mail.nwpu.edu.cn, \{shaanyang,dansu\}@tencent.com \\ lxie@nwpu.edu.cn}

\begin{document}
\begin{CJK*}{UTF8}{gbsn}

\maketitle
\begin{abstract}
Current two-stage TTS framework typically integrates an acoustic model with a vocoder -- the acoustic model predicts a low resolution intermediate representation such as Mel-spectrum while the vocoder generates waveform from the intermediate representation. Although the intermediate representation is served as a bridge, there still exists critical mismatch between the acoustic model and the vocoder as they are commonly separately learned and work on different distributions of representation, leading to inevitable artifacts in the synthesized speech. In this work, different from using pre-designed intermediate representation in most previous studies, we propose to use VAE combining with GAN to learn a latent representation directly from speech and then utilize a flow-based acoustic model to model the distribution of the latent representation from text. In this way, the mismatch problem is migrated as the two stages work on the same distribution. Results demonstrate that the flow-based acoustic model can exactly model the distribution of our learned speech representation and the proposed TTS framework, namely Glow-WaveGAN, can produce high fidelity speech outperforming the state-of-the-art GAN-based model.
\end{abstract}


\section{Introduction}
With the continuous development of deep learning, text-to-speech (TTS) has made tremendous progress. Due to the huge difference in temporal resolution between text and audio samples, most approaches divide the TTS procedure into two stages, namely acoustic model and vocoder. The acoustic model first generates frame-level intermediate representations, such as Mel-spectrum given character or phoneme sequences, while the vocoder produces audio samples conditioned on the intermediate representation.

Attention-based auto-regressive (AR) model, such as Tacotron~\cite{wang2017tacotron} and Tacotron2~\cite{shen2018natural}, is a typical framework of acoustic modeling, which takes character or phoneme sequences as input and generates Mel-spectrum frame by frame. The AR models have achieved reasonably naturalness with human-parity speech in closed domains. The other family is the non-AR acoustic model that can synthesize speech much faster in a parallel manner. The key to design a non-AR acoustic model is alignment prediction. FastSpeech~\cite{fastspeech} was proposed to learn alignment from pre-trained AR teacher model, while FastSpeech2~\cite{ren2021fastspeech} bypassed the requirement of knowledge from the teacher model through an external force aligner to extract duration label. Glow-TTS~\cite{Glow-TTS} was a flow-based generative model for parallel generation, which obtains the alignment by the properties of flows and dynamic programming.

For the vocoder that generates audio samples from intermediate representation, most high-fidelity neural vocoders are auto-regressive~\cite{oord2016wavenet,kalchbrenner2018efficient,mehri2016samplernn}, which produce outputs relying on the previous sample. However, because of the AR structure, it's prohibitively slow in synthesizing high temporal resolution audio. The neural vocoders~\cite{kumar2019melgan, yamamoto2020parallel, jang2020universal, yang2020multi} based on GAN (generative adversarial networks) are proposed to address this problem. They all rely on an adversarial game of two networks, a generator network that produces samples conditioned with the intermediate speech representation and a discriminator or several discriminators to differentiate between real and generated samples. And to stabilize the training process, many studies~\cite{kumar2019melgan, yamamoto2020parallel, yang2020multi} adopted auxiliary loss, i.e. multi-resolution STFT loss and feature matching loss.

Despite the recent success on acoustic models and neural vocoders, there still exists critical mismatch between the two-stages, which may cause artifacts in the synthesized speech during inference. We believe a decisive reason is that the generated acoustic features from text lie in a different distribution with those extracted directly from speech. Note that the conventional acoustic models often use mean square error (MSE) to estimate model parameters, which only considers the numerical differences and may lead to over-smoothing problems~\cite{kaneko2017generative}. To overcome the above issue, generative adversarial networks are utilized to model the distribution of target speech~\cite{kaneko2017generative,yang2017statistical}. Another intuitive way to alleviate the mismatch is to fine-tune the neural vocoder with the generated speech representation, which is widely used in the practical sequence-to-sequence based acoustic models~\cite{shen2018natural,donahue2020end}.  But in the auto-regressive manner, the ground-truth alignment mode is still different from the inference mode. Integrating acoustic model and vocoder into an end-to-end model is another possible solution~\cite{ren2021fastspeech, donahue2020end, weiss2020wave, miao2020efficienttts}, but this kind of models are notoriously difficult to train and extra Mel-spectrum based loss has to be adopted to stabilize the convergence process.

In our work, we propose Glow-WaveGAN, which utilizes a flow-based acoustic model to learn the distribution of hidden representations from a variational auto-encoder (VAE)~\cite{kingma2013auto}. Different from all the existing methods, we do not directly use any predesigned acoustic features such as Mel-spectrum. This is because the estimated approximations of the Mel distribution from the acoustic model are still different from those inputs to a neural vocoder, which may cause artifacts i.e., metallic effects in synthesized audios. To make sure the target distribution of the acoustic model is exactly from the modeled distribution from the vocoder, we propose to learn a hidden representation directly from speech through variational auto-encoder (VAE)~\cite{kingma2013auto}, where we further apply the generative adversarial network (GAN)~\cite{goodfellow2014generative} on the outputs to improve the quality of reconstructed speech.

Specifically, our proposed Glow-WaveGAN consists of a WaveGAN and a Flow-based acoustic model. The proposed WaveGAN utilizes GAN-based variational auto-encoder to learn speech latent representation. The encoder in VAE acts as an extractor to extract speech representation, while the decoder works as a vocoder to reconstruct waveform from the latent representation. With the extracted latent representation from speech via the well-trained WaveGAN, we then employ a flow-based acoustic model to learn the distribution of the latent representation from text. Experiments show that our proposed approach can produce high-fidelity speech, obviously outperforming the popular GAN-based models in audio quality.


\section{Proposed method}
\begin{figure}[htb]
  \centering
  \setlength{\belowcaptionskip}{3pt}
  \setlength{\abovecaptionskip}{3pt}
  \includegraphics[width=0.9\linewidth]{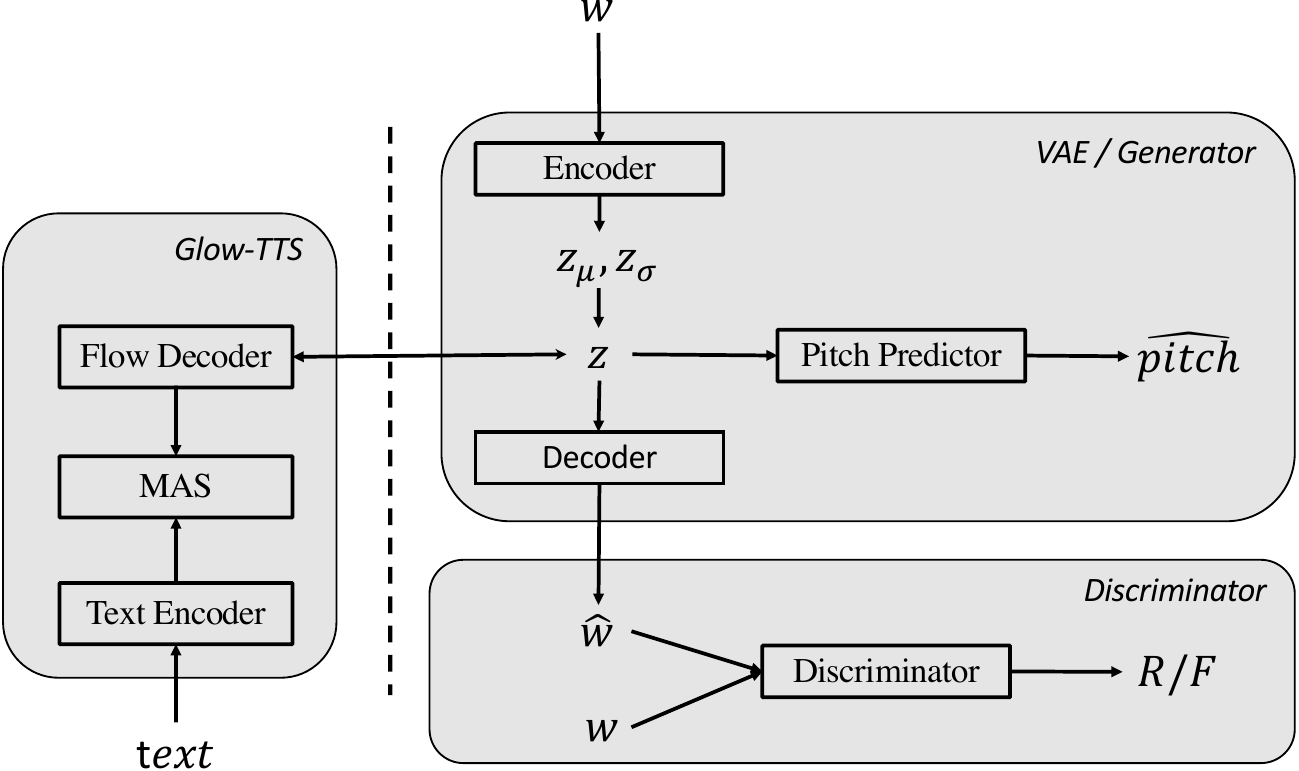}
  \caption{Model architecture of proposed Glow-WaveGAN.}
  \label{fig:am_architecture}
\end{figure}

Fig.~\ref{fig:am_architecture} illustrates the proposed Glow-WaveGAN for high-fidelity speech synthesis. The architecture consists of a WaveGAN module and a Glow-TTS~\cite{Glow-TTS} module. The WaveGAN module is intended to learn latent speech representation by conducting waveform reconstruction, while the Glow-TTS model is designed to map input text to speech representation learned by the WaveGAN module. To learn a robust distribution of the latent speech representation, we utilize a variational auto-encoder (VAE)~\cite{kingma2013auto} to conduct feature extraction and reconstruction. To improve the quality of reconstructed speech, we further introduce pitch prediction as an auxiliary task with the latent representation as input and then employ
the generative adversarial network~\cite{goodfellow2014generative} to model the distribution of target speech. Finally, we utilize the flow-based acoustic model to learn the distribution of the latent representation extracted from the encoder of the VAE. In this way, the main idea of our proposed Glow-WaveGAN is to make both the acoustic model and the waveform reconstruction module follow the same distribution, which successfully produces high quality natural speech compared to the state-of-the-art models.

\subsection{VAE for learning representation and reconstruction}\label{sec:vae}
To relieve the mismatch between the acoustic model and the vocoder in conventional two-stage speech synthesis, we propose to force the acoustic model to directly learn the distribution of a latent feature from the vocoder. To achieve this goal, we first use VAE to learn speech representation in an unsupervised manner. Conventional VAE typically contains an encoder and a decoder, where the encoder encodes the inputs $w$ into hidden features $z$ and the decoder tries to reconstruct $w$ from $z$:
\begin{equation}
  \setlength{\abovedisplayskip}{3pt}
  \setlength{\belowdisplayskip}{3pt}
  z = Enc(w) \sim q(z|w)
  \label{eq_encoder}
\end{equation}
\begin{equation}
  \setlength{\abovedisplayskip}{3pt}
  \setlength{\belowdisplayskip}{3pt}
  \hat{w} = Dec(z) \sim p(w|z)
  \label{eq_2}
\end{equation}
where $w$ means the input waveform. $q(z|w)$ is the distribution of hidden features, which is utilized to reconstruct waveform $\hat w$ that lies in $p(w|z)$.

Conventional VAE usually contains a reconstruction objective $L_{recon}$ and a prior regularization term as
\begin{equation}
  \setlength{\abovedisplayskip}{3pt}
  \setlength{\belowdisplayskip}{3pt}
  L_{vae} = L_{recon} + D_{KL}(q(z|w)||p(z)),
  \label{eq_vae_loss}
\end{equation}
where $D_{KL}$ is the Kullback-Leibler divergence and $p(z)$ is the prior distribution. As for the proposed VAE for speech reconstruction, we adopt multi-resolution STFT loss~\cite{yamamoto2020parallel, binkowski2019high,yamamoto2019probability} to measure the distance between natural speech $w$ and generated speech $\hat{w}$ in frequency domain. 
Carrying out the multiple STFT losses with different analysis parameters is proved to be effective to learn the time-frequency characteristics of speech~\cite{kumar2019melgan}. 

\begin{figure}[htb]
  \centering
  \includegraphics[width=0.7\linewidth]{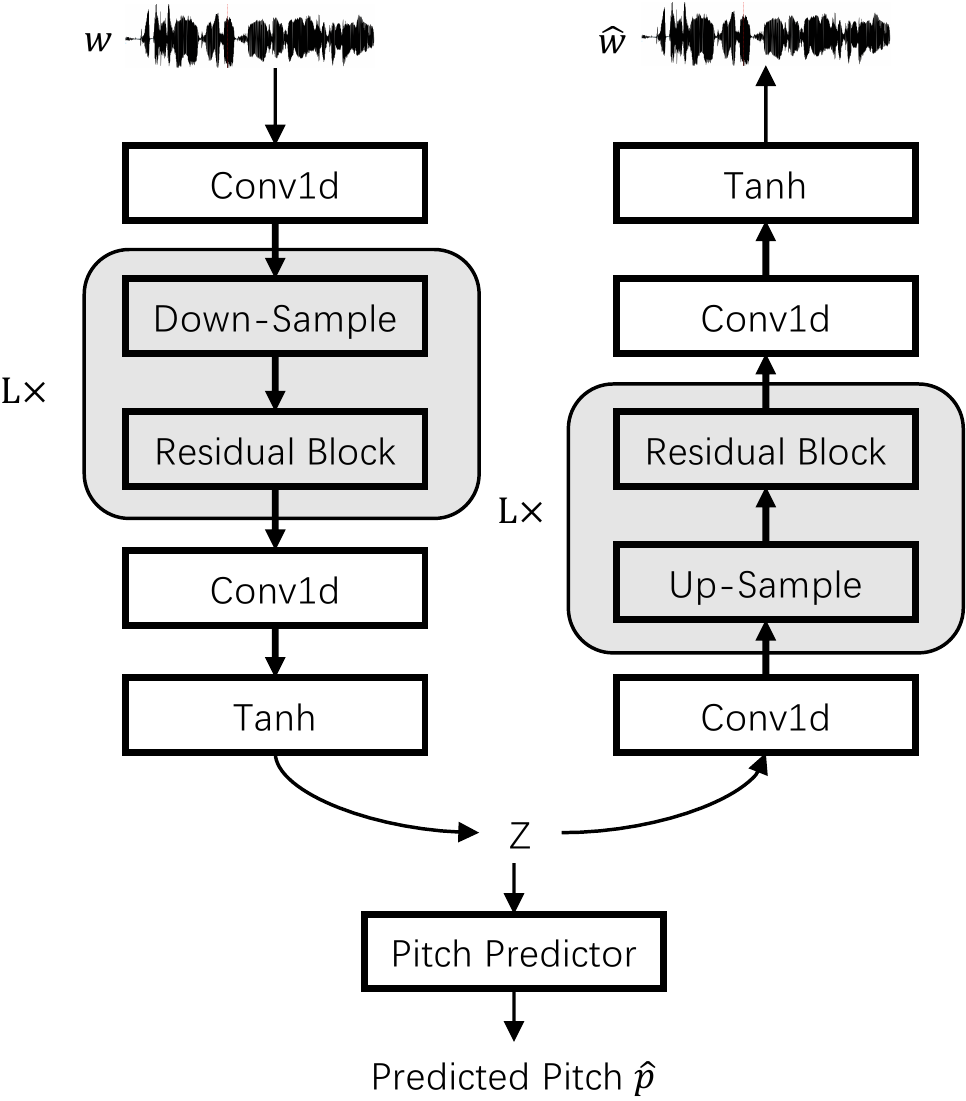}
  \caption{Model architecture of proposed auto-encoder for learning hidden features.}
  \label{fig:ae-WaveGAN}
\end{figure}

The detailed architecture of the proposed VAE for learning speech representation is shown in Fig.~\ref{fig:ae-WaveGAN}. The encoder contains a stack of down-sampling convolution layers followed by a residual block to summarize the abstract information in speech. Since it is hard for the acoustic model to learn long-time dependency in sample-level, in the residual block, we adopt the dilated convolutions to increase the receptive field to model the long-range context~\cite{kumar2019melgan}. After the encoding process, we treat the produced mean and variance as the statistics of the learned latent distribution $q(z|w)=\mathcal{N}(Z_\mu, Z_\sigma)$. The structure of the decoder is mirror symmetry of the encoder, except that we employ a stack of transposed convolution with residual blocks to up-sample the sampled $z$ to speech $\hat{w}$.

Since pitch information is important for natural speech synthesis and $z$ is treated as the target of the acoustic model, we further propose to enhance the latent representation $z$ with pitch information. To achieve this, we introduce a pitch predictor to predict pitch information from the sampled $z$ with a frame-level pitch reconstruction objective as:
\begin{equation}
  \setlength{\abovedisplayskip}{3pt}
  \setlength{\belowdisplayskip}{3pt}
  L_{pitch} = \Vert f(z) - p \Vert_2
  \label{eq:pitch_loss}
\end{equation}
where $p$ represents the extracted pitch in log-scale, $f(\cdot)$ is the predictor with two convolution layers followed by a linear output layer. The frame-shift in pitch extraction depends on the total down-sampling factor.

\subsection{GAN for high-fidelity speech reconstruction}
Since speech reconstructed by the above auto-encoder may exist audio artifacts, we further propose to use GAN to improve the quality of the reconstructed speech, in viewing of the recent success of GAN-based vocoders~\cite{kumar2019melgan,yamamoto2020parallel,kong2020hifi}. As shown in Fig.~\ref{fig:am_architecture}, we treat the above VAE with pitch predictor as the generator of GAN. Hence, the generator firstly encodes the waveform into hidden $z$, and then produces speech from sampled $z$ of $q(z|w)$. The discriminator network is optimized to classify the real samples $w$ and the generated samples $\hat{w}$. During synthesis, only the decoder part of the generator is used to reconstruct waveform.

As for the adversarial training, we utilize the least-squares GANs objectives~\cite{mao2017least} to stabilize the training. The GAN losses for generator and discriminator are defined as:
\begin{equation}
  \setlength{\abovedisplayskip}{3pt}
  \setlength{\belowdisplayskip}{3pt}
  L_{adv\_g} = (D(G(w)) - 1)^2
  \label{eq_decoder}
\end{equation}
\begin{equation}
  \setlength{\abovedisplayskip}{3pt}
  \setlength{\belowdisplayskip}{3pt}
  L_{adv\_d} = (D(w) - 1)^2 + D(G(w))^2
  \label{eq_decoder}
\end{equation}
where $G$ represents both the encoding and decoding parts in the VAE. $D$ denotes multi-resolution spectrum discriminator~\cite{jang2020universal} operating on Mel-spectrum with different analysis parameters. 

Furthermore, we use the feature matching loss~\cite{kumar2019melgan} as an additional loss to train the generator. The feature matching loss aims to minimize the L1 distance between the feature map extracted from intermediate layers in each discriminator, which is defined as:
\begin{equation}
  \setlength{\abovedisplayskip}{3pt}
  \setlength{\belowdisplayskip}{3pt}
  L_{fm} = \sum_{s}\sum_{i}{\frac{1}{N_i}\Vert D_s^i(w) - D_s^i(G(w)) \Vert_1},
  \label{feature_match}
\end{equation}
where $D_s^i(\cdot)$ is the output from the $i^{th}$ in $s^{th}$ discriminator, and $N_i$ denotes the number of units in each layer.

With the above VAE with pitch predictor and the GAN, we optimize the WaveGAN with the full objective
\begin{equation}
  \setlength{\abovedisplayskip}{3pt}
  \setlength{\belowdisplayskip}{3pt}
  L = \lambda_1L_{kl} + \lambda_2L_{pitch} + \lambda_3L_{recons} + \lambda_4L_{adv\_g} + \lambda_5L_{fm}
  \label{loss_whole}
\end{equation}
where the $L_{kl}$ corresponds the Kullback-Leibler divergence in~Eq.(\ref{eq_vae_loss}) and we empirically set $\lambda_1$=10, $\lambda_2$=1, $\lambda_3$=1, $\lambda_4$=1, $\lambda_5$=20. The discriminator is optimized with adversarial loss $L_{adv\_d}$. The whole training procedure is summarized in Algorithm~\ref{alg:Framwork}. 

\begin{algorithm}[htb]
  \footnotesize
  \caption{Training the WaveGAN model}  
  \label{alg:Framwork}  
  \begin{algorithmic}
    \State $\{\Theta_E, \Theta_G, \Theta_D, \Theta_{PN}\}\leftarrow$ initial encoder, decoder  (generator) discriminator and pitch predictor network parameters
    \Repeat
    \State $W\leftarrow$ random mini-batch audios from data-set
    \State $P\leftarrow$ pitch corresponding to $W$
    \State $\{Z_{\mu}, Z_{\sigma}\}\leftarrow$ E(W)
    \State $Z\leftarrow$ sample from $\mathcal{N}(Z_{\mu}, Z_{\sigma})$
    \State $L_{kl}\leftarrow KL(P_z\Vert\mathcal{N}(0, I))$
    \State $\hat{P}\leftarrow f(Z)$ // predict pitch using pitch predictor 
    \State $\hat{W}\leftarrow G(Z)$
    \State $L_{pitch}\leftarrow\Vert \hat{P} - P \Vert_2$
    \State $L_{recons}\leftarrow$ multi-resolution STFT loss.
    \State $L_{adv\_g}\leftarrow (D(\hat{W}) - 1)^2$
    \State $L_{adv\_d}\leftarrow(D(W) - 1)^2 + D(\hat{W})^2$
    \State $L_{fm}\leftarrow$ feature matching loss with~Eq.~(\ref{feature_match})
    \State // Update parameters according to gradients.
    \State $\Theta_E\stackrel{+}{\leftarrow}-\nabla_{\Theta_E}(L_{kl}+L_{pitch}+L_{fm} + L_{adv\_g} + L_{recons})$
    \State $\Theta_P\stackrel{+}{\leftarrow}-\nabla_{\Theta_P}(L_{pitch})$
    \State $\Theta_G\stackrel{+}{\leftarrow}-\nabla_{\Theta_G}(L_{fm} + L_{adv\_g} + L_{recons}))$
    \State $\Theta_D\stackrel{+}{\leftarrow}-\nabla_{\Theta_D}(L_{adv\_d})$
    \Until{done}
  \end{algorithmic}  
\end{algorithm}
Note that the architecture only combining the decoder VAE and the discriminator can be treated as a GAN-based vocoder like~\cite{kumar2019melgan, kong2020hifi} when replacing the speech representation $Z$ with Mel-spectrum. And we call the combination vocoder \textit{Inner-GAN} for experimental comparison.

\subsection{Flow based acoustic model}
With the WaveGAN including VAE and GAN, we can obtain an extractor to extract latent representations and a speech reconstructor to synthesize waveform from the representations. Accordingly, acoustic model is designed to model the distribution of the latent represent $z$. To achieve this goal, we employ the flow-based acoustic model -- Glow-TTS~\cite{Glow-TTS} to directly maximize the likelihood of $z$ by applying invertible transformations.

Firstly, we extract the statistics $Z_{\mu}, Z_{\sigma}$ of the training data. Hence, the Glow-TTS models the conditional distribution of speech representation $P_Z(z|t)$ by transforming a conditional prior distribution $P_C(c|t)$ through the flow-based decoder $f_{dec}:c \rightarrow z$, where $t$ denotes the text sequences. As for the prior distribution $P_C(c|t)$, we compute its statistics $C_{\mu}=C_{\mu_{1:M}}$ and $C_{\sigma}=C_{\sigma_{1:M}}$ from each token in the text $t=t_{1:M}$, where $M$ is the sequence length.

We can calculate the log-likelihood of the $Z$ using change of variables as follows:
\begin{equation}
  \setlength{\abovedisplayskip}{3pt}
  \setlength{\belowdisplayskip}{3pt}
  \log P_Z(z|t) = logP_C(c|t) + \log \vert det \frac{\partial f^{-1}_{dec}(z)}{\partial_z } \vert
  \label{loss_whole}
\end{equation}
where the prior distribution $P_C$ follows an isotropic multivariate Gaussian distribution.

Since the sequence of target variable $z$ and the input variable $t$ are totally different, we need to align these two sequences to compute $logP_C(c|t)$. Same as Glow-TTS~\cite{Glow-TTS}, we also use the Monotonic Alignment Search (MAS) to map from $z$ to each $c$, where $A(j)=i$ if $c_j \sim \mathcal{N}(c_j; C_{\mu_i}, C_{\sigma_i})$. So the log-likelihood of prior distribution is:
\begin{equation}
  \setlength{\abovedisplayskip}{3pt}
  \setlength{\belowdisplayskip}{3pt}
  logP_C(c|t;\Theta,A) = \sum_{j=1}^{N}log\mathcal{N}(c_j; C_{\mu_{A(j)}},C_{\sigma_{A(j)}}).
  \label{loss_whole}
\end{equation}
where $N$ is the length of the speech representation $Z$ from the above WaveGAN. The trainable variables in parameter set $\Theta$ and the alignment $A$ are optimized iteratively~\cite{Glow-TTS}. 

With the Glow-TTS, we transform the explicit distribution $\mathcal{N}(Z_\mu, Z_\sigma)$ from WaveGAN
into the prior conditional distribution $P(c|t)$ during training. At each training step, we sample the target $z$ from $\mathcal{N}(Z_{\mu}, Z_{\sigma})$ instead of using pre-sampled constant targets. At inference time, the flow-based acoustic model can produce $\hat z$ that follows the distribution $\mathcal{N}(Z_\mu, Z_\sigma)$ from texts, which is further fed into the decoder part of the WaveGAN to reconstruct high-fidelity synthesized speech.

\section{Experiments and Results}
\subsection{Basic setups}
To evaluate the proposed method, we conduct experiments on two English datasets. For the single speaker setting, we use LJSpeech~\cite{ljspeech17}, which consists of 13,100 audio clips in 22.5 kHz with a total duration of approximately 24 hours. We randomly reserve 500 samples as test set.
For the multi-speaker setting, we use VCTK~\cite{veaux2016superseded} corpus, which consists of approximately 44,200 audio clips uttered by 109 native English speakers with various accents. The total length of the audio clips is approximately 44 hours and the sample rate is 44kHz. We reduce the sample rate to 24kHz and randomly select nine speakers for testing and exclude all their audio clips from the training set\footnote{Samples can be found at \url{https://syang1993.github.io/glow_wavegan}}. We evaluate our proposed approach on speech synthesis by training several models on the datasets described above:
\begin{itemize}
  \item [1)] 
  The state of the art GAN-based vocoder \textit{HiFi-GAN}~\cite{kong2020hifi}. 
  \item [2)]
  The \textit{Inner-GAN} vocoder which combines the VAE decoder and the discriminator of our proposed model mentioned in Section~\ref{sec:vae}
  \item [3)]
  Our proposed \textit{Glow-WaveGAN}. 
 \end{itemize}
 We use the publicly available implementation of HiFi-GAN vocoder~\footnote{\url{https://github.com/jik876/HiFi-GAN}}. The same original implementation of Glow-TTS\footnote{\url{https://github.com/jaywalnut310/glow-tts}} is adopted for all our experiments, and phonemes are chosen as input text tokens. For our proposed Glow-WaveGAN, the encoder down-samples audio to $Z$ by a factor of 256 and the dimension of $Z$ is 256. The down-sample factors and the corresponding channels are [2,4,4,8] and [128, 128, 256, 512]. The decoder takes 256 dimension $Z$ as input and up-samples 256 factors to audio space, while the up-sample factors and channels are [8,4,4,2] and [256, 128, 128, 64]. For the discriminator, we use six different spectrum discriminators.
\newcommand{\tabincell}[2]{\begin{tabular}{@{}#1@{}}#2\end{tabular}}  
\subsection{Evaluation on text-to-speech}
We conduct the opinion score (MOS) tests in terms of naturalness and audio quality on the test set of LJSpeech and VCTK. The MOS scores are listed in Table~\ref{table:t1}. The results show whether reconstruction to waveform (copysyn) or text to speech, the proposed Glow-WaveGAN scores consistently higher than the other models. And the gap between the text to speech and the reconstruction to waveform of the proposed Glow-WaveGAN is only 0.08, which is significantly lower than that of HiFi-GAN (0.6). It implies that the speech representation with exact distribution has learned the true distribution of speech and the learned distribution is easier for the acoustic model to estimate. And the results of Inner-GAN further verifies that the improvement from the proposed Glow-WaveGAN compared with HiFi-GAN is not due to the difference between the model architecture or auxiliary losses. For the comparison between Inner-GAN and HiFi-GAN, the Inner-GAN scores lower than HiFi-GAN. We believe that the gap is because that the HiFi-GAN model adopts the multi-receptive field fusion module and takes more discriminators. 
\begin{table}[!htb]
  \small
  \centering
  \setlength{\belowcaptionskip}{-4.5pt}
  \caption[]{The MOS results of a multi-speaker TTS on VCTK and single-speaker TTS on LJSpeech. SR stands for speech representation ($Mel$ and $Z$ are extracted directly from speech, while $\hat{Mel}$ and $\hat{Z}$ are predicted from acoustic model).}
  \resizebox{0.75\linewidth}{!}{
    \begin{tabular}{l|c|cc}
      \hline \hline
      Model & SR & VCTK & LJ \\ \hline \hline
      Ground Truth & &4.5 & 4.6 \\ \hline \hline
      HiFi-GAN & $Mel$ & 3.8 & 3.86 \\ \hline    
      Glow + HiFi-GAN & $\hat{Mel}$ & 3.21 &  3.4 \\ \hline
      Inner-GAN & $Mel$ & 3.78 & 3.84 \\ \hline
      Glow + Inner-GAN & $\hat{Mel}$ & 3.20 & 3.35 \\ \hline \hline
      WaveGAN & $Z$ & \textbf{4.13} & \textbf{4.15} \\ \hline
      Glow-WaveGAN &  $\hat{Z}$ & \textbf{4.05} & \textbf{4.09} \\ \hline 
      \tabincell{l}{Glow-WaveGAN \\ (w/o pitch predictor)} & $\hat{Z}$ & 3.87 & 3.93 \\ \hline \hline
      \end{tabular}
  }
  \label{table:t1}
\end{table}

\begin{figure}[!htb]
  \centering
  \includegraphics[width=0.78\linewidth]{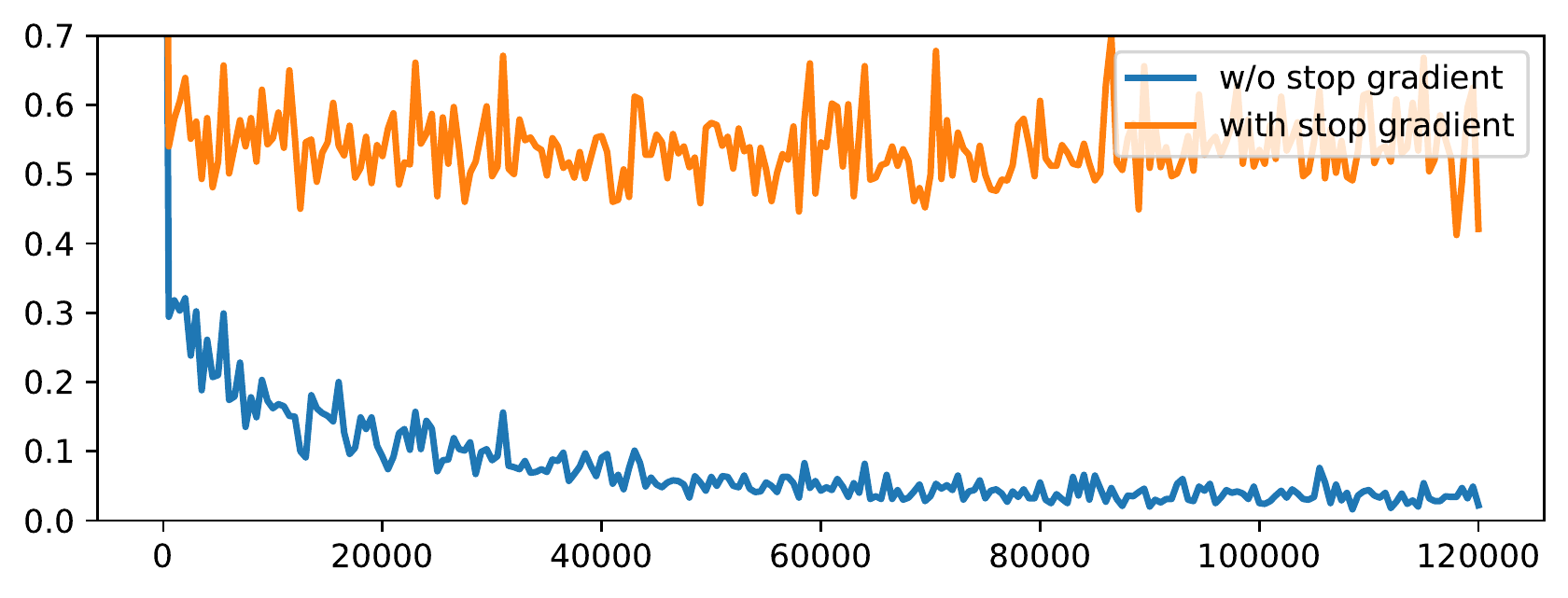}
  \caption{The pitch loss of our proposed model at the training procedure.}
  \label{fig:pitch_loss}
\end{figure}

The ablation without pitch prediction achieved substantially worse MOS, demonstrating the importance of adding pitch supervision in learning the speech representation $Z$. To further examine the effectiveness of the pitch predictor, we use the speech representation $Z$ without pitch supervision to predict pitch. Specifically, we train Glow-WaveGAN and apply stop gradient operator to the input of the pitch predictor. The pitch loss curve is shown in Fig.~\ref{fig:pitch_loss}. We can see that the pitch loss of the model with stop gradient even does not get converge, which demonstrates that there is no exact pitch information in the speech representation $Z$ if we do not utilize the pitch predictor. As a result, the listeners point out many jitters in the speech synthesized by Glow-WaveGAN without the pitch predictor.

\subsection{Generalization to unseen speakers}
To measure the generalization ability of our model to unseen speakers,  we use 100 randomly selected utterances from nine unseen speakers in the VCTK for MOS test. Note that for training of vocoder and WaveGAN, the utterances of unseen speakers are totally excluded from the training set, but for the training of Glow-TTS, only the test set for MOS test is excluded. 

The results are shown in Table~\ref{table:unseen}. We can see that our proposed model scores better than other models, and the score is even close to that of seen speakers, indicating that the proposed model generalizes well to unseen speakers. For the HiFi-GAN and Inner-GAN, due to the mismatch between the predicted and the ground truth Mel-spectrum, listeners can tell more artifacts like metallic sounds from the synthesized samples.
\begin{table}[!htb]
  \small
  \centering
  \setlength{\belowcaptionskip}{-5pt}
  \setlength{\abovecaptionskip}{3pt}
  \caption[]{The MOS results for the unseen speaker in VCTK.}
  \resizebox{0.72\linewidth}{!}{
    \begin{tabular}{l|c|c}
      \hline \hline
      Model & SR & MOS \\ \hline \hline
      Ground Truth &  & ~~~~~~~4.5~~~~~~~ \\ \hline
      HiFi-GAN & $Mel$ & 3.76 \\ \hline
      Glow + HiFi-GAN & $\hat{Mel}$ & 3.14 \\ \hline \hline
      Inner-GAN  & $Mel$ & 3.70 \\ \hline
      Glow + Inner-GAN  & $\hat{Mel}$ & 3.12 \\ \hline
      WaveGAN & $Z$ & \textbf{4.08} \\ \hline
      Glow-WaveGAN & $\hat{Z}$ & \textbf{4.0} \\ \hline \hline
      \end{tabular}
  }
  \label{table:unseen}
\end{table}
\vspace{-0.5cm}
\section{Conclusions}
In this work, we introduce Glow-WaveGAN, which can synthesize high fidelity speech from text, without using Mel-spectrum as the intermediate representation. Specifically, we propose WaveGAN which combines VAE and GAN to extract a latent speech representation by encoder and the representation can be reconstructed back to waveform by decoder. And to model the distribution of latent representation, we utilize Glow-TTS to estimate the distribution from text. Through this way, we migrate the problematic data distribution mismatch between acoustic model and vocoder in a typical two-stage TTS system. Results show that our proposed model outperforms the state-of-the-art GAN-based text-to-speech framework in audio quality.

\bibliographystyle{IEEEtran}
\bibliography{template}

\begin{thebibliography}{10}
\providecommand{\url}[1]{#1}
\csname url@samestyle\endcsname
\providecommand{\newblock}{\relax}
\providecommand{\bibinfo}[2]{#2}
\providecommand{\BIBentrySTDinterwordspacing}{\spaceskip=0pt\relax}
\providecommand{\BIBentryALTinterwordstretchfactor}{4}
\providecommand{\BIBentryALTinterwordspacing}{\spaceskip=\fontdimen2\font plus
\BIBentryALTinterwordstretchfactor\fontdimen3\font minus
  \fontdimen4\font\relax}
\providecommand{\BIBforeignlanguage}[2]{{%
\expandafter\ifx\csname l@#1\endcsname\relax
\typeout{** WARNING: IEEEtran.bst: No hyphenation pattern has been}%
\typeout{** loaded for the language `#1'. Using the pattern for}%
\typeout{** the default language instead.}%
\else
\language=\csname l@#1\endcsname
\fi
#2}}
\providecommand{\BIBdecl}{\relax}
\BIBdecl

\bibitem{wang2017tacotron}
Y.~Wang, R.~J. Skerry{-}Ryan, D.~Stanton, Y.~Wu, R.~J. Weiss, N.~Jaitly,
  Z.~Yang, Y.~Xiao, Z.~Chen, S.~Bengio, Q.~V. Le, Y.~Agiomyrgiannakis,
  R.~Clark, and R.~A. Saurous, ``Tacotron: Towards end-to-end speech
  synthesis,'' in \emph{Proc. of INTERSPEECH}, 2017, pp. 4006--4010.

\bibitem{shen2018natural}
J.~Shen, R.~Pang, R.~J. Weiss, M.~Schuster, N.~Jaitly, Z.~Yang, Z.~Chen,
  Y.~Zhang, Y.~Wang, R.~Ryan, R.~A. Saurous, Y.~Agiomyrgiannakis, and Y.~Wu,
  ``Natural {TTS} synthesis by conditioning wavenet on {MEL} spectrogram
  predictions,'' in \emph{Proc. of ICASSP}, 2018, pp. 4779--4783.

\bibitem{fastspeech}
Y.~Ren, Y.~Ruan, X.~Tan, T.~Qin, S.~Zhao, Z.~Zhao, and T.~Liu, ``Fastspeech:
  Fast, robust and controllable text to speech,'' in \emph{Proc. of NeurIPS},
  2019, pp. 3165--3174.

\bibitem{ren2021fastspeech}
Y.~Ren, C.~Hu, X.~Tan, T.~Qin, S.~Zhao, Z.~Zhao, and T.-Y. Liu, ``Fastspeech 2:
  Fast and high-quality end-to-end text to speech,'' in \emph{Proc. of ICLR},
  2021.

\bibitem{Glow-TTS}
J.~Kim, S.~Kim, J.~Kong, and S.~Yoon, ``Glow-tts: {A} generative flow for
  text-to-speech via monotonic alignment search,'' in \emph{Proc. of NeurIPS},
  2020.

\bibitem{oord2016wavenet}
A.~van~den Oord, S.~Dieleman, H.~Zen, K.~Simonyan, O.~Vinyals, A.~Graves,
  N.~Kalchbrenner, A.~Senior, and K.~Kavukcuoglu, ``Wavenet: A generative model
  for raw audio,'' in \emph{9th ISCA Speech Synthesis Workshop}, 2016, pp.
  125--125.

\bibitem{kalchbrenner2018efficient}
N.~Kalchbrenner, E.~Elsen, K.~Simonyan, S.~Noury, N.~Casagrande, E.~Lockhart,
  F.~Stimberg, A.~van~den Oord, S.~Dieleman, and K.~Kavukcuoglu, ``Efficient
  neural audio synthesis,'' in \emph{Proc. of ICML}, 2018, pp. 2415--2424.

\bibitem{mehri2016samplernn}
S.~Mehri, K.~Kumar, I.~Gulrajani, R.~Kumar, S.~Jain, J.~Sotelo, A.~C.
  Courville, and Y.~Bengio, ``Samplernn: An unconditional end-to-end neural
  audio generation model,'' in \emph{Proc. of ICLR}, 2017.

\bibitem{kumar2019melgan}
K.~Kumar, R.~Kumar, T.~de~Boissiere, L.~Gestin, W.~Z. Teoh, J.~Sotelo,
  A.~de~Br{\'{e}}bisson, Y.~Bengio, and A.~C. Courville, ``Melgan: Generative
  adversarial networks for conditional waveform synthesis,'' in \emph{Proc. of
  NeurIPS}, 2019, pp. 14\,881--14\,892.

\bibitem{yamamoto2020parallel}
R.~Yamamoto, E.~Song, and J.~Kim, ``Parallel wavegan: {A} fast waveform
  generation model based on generative adversarial networks with
  multi-resolution spectrogram,'' in \emph{Proc. of ICASSP}, 2020, pp.
  6199--6203.

\bibitem{jang2020universal}
W.~Jang, D.~Lim, and J.~Yoon, ``Universal melgan: A robust neural vocoder for
  high-fidelity waveform generation in multiple domains,'' \emph{arXiv preprint
  arXiv:2011.09631}, 2020.

\bibitem{yang2020multi}
G.~Yang, S.~Yang, K.~Liu, P.~Fang, W.~Chen, and L.~Xie, ``Multi-band melgan:
  Faster waveform generation for high-quality text-to-speech,'' in \emph{{IEEE}
  Spoken Language Technology Workshop, {SLT} 2021, Shenzhen, China, January
  19-22, 2021}, 2021, pp. 492--498.

\bibitem{kaneko2017generative}
T.~Kaneko, H.~Kameoka, N.~Hojo, Y.~Ijima, K.~Hiramatsu, and K.~Kashino,
  ``Generative adversarial network-based postfilter for statistical parametric
  speech synthesis,'' in \emph{Proc. of ICASSP}, 2017, pp. 4910--4914.

\bibitem{yang2017statistical}
S.~Yang, L.~Xie, X.~Chen, X.~Lou, X.~Zhu, D.~Huang, and H.~Li, ``Statistical
  parametric speech synthesis using generative adversarial networks under a
  multi-task learning framework,'' in \emph{2017 IEEE Automatic Speech
  Recognition and Understanding Workshop (ASRU)}, 2017, pp. 685--691.

\bibitem{donahue2020end}
J.~Donahue, S.~Dieleman, M.~Bi{\'n}kowski, E.~Elsen, and K.~Simonyan,
  ``End-to-end adversarial text-to-speech,'' \emph{arXiv preprint
  arXiv:2006.03575}, 2020.

\bibitem{weiss2020wave}
R.~J. Weiss, R.~Skerry-Ryan, E.~Battenberg, S.~Mariooryad, and D.~P. Kingma,
  ``Wave-tacotron: Spectrogram-free end-to-end text-to-speech synthesis,'' in
  \emph{Proc. of ICASSP}, 2021, pp. 5679--5683.

\bibitem{miao2020efficienttts}
C.~Miao, S.~Liang, Z.~Liu, M.~Chen, J.~Ma, S.~Wang, and J.~Xiao,
  ``Efficienttts: An efficient and high-quality text-to-speech architecture,''
  \emph{arXiv preprint arXiv:2012.03500}, 2020.

\bibitem{kingma2013auto}
D.~P. Kingma and M.~Welling, ``Auto-encoding variational bayes,'' in
  \emph{Proc. of ICLR}, 2014.

\bibitem{goodfellow2014generative}
I.~J. Goodfellow, J.~Pouget{-}Abadie, M.~Mirza, B.~Xu, D.~Warde{-}Farley,
  S.~Ozair, A.~C. Courville, and Y.~Bengio, ``Generative adversarial nets,'' in
  \emph{Proc. of NeurIPS}, 2014, pp. 2672--2680.

\bibitem{binkowski2019high}
M.~Binkowski, J.~Donahue, S.~Dieleman, A.~Clark, E.~Elsen, N.~Casagrande, L.~C.
  Cobo, and K.~Simonyan, ``High fidelity speech synthesis with adversarial
  networks,'' in \emph{Proc. of ICLR}, 2020.

\bibitem{yamamoto2019probability}
R.~Yamamoto, E.~Song, and J.~Kim, ``Probability density distillation with
  generative adversarial networks for high-quality parallel waveform
  generation,'' in \emph{Proc. of INTERSPEECH}, 2019, pp. 699--703.

\bibitem{kong2020hifi}
J.~Kong, J.~Kim, and J.~Bae, ``Hifi-gan: Generative adversarial networks for
  efficient and high fidelity speech synthesis,'' in \emph{Proc. of NeurIPS},
  2020.

\bibitem{mao2017least}
X.~Mao, Q.~Li, H.~Xie, R.~Y.~K. Lau, Z.~Wang, and S.~P. Smolley, ``Least
  squares generative adversarial networks,'' in \emph{Proc. of ICCV}, 2017, pp.
  2813--2821.

\bibitem{ljspeech17}
K.~Ito and L.~Johnson, ``The lj speech dataset,''
  \url{https://keithito.com/LJ-Speech-Dataset/}, 2017.

\bibitem{veaux2016superseded}
C.~Veaux, J.~Yamagishi, K.~MacDonald \emph{et~al.}, ``Superseded-cstr vctk
  corpus: English multi-speaker corpus for cstr voice cloning toolkit,'' 2016.

\end{thebibliography}
\end{CJK*}
\end{document}